\documentclass[amsmath,superscriptaddress,amssymb,aps,twocolumn]{revtex4}
\usepackage{graphicx}
\usepackage{soul}
\usepackage[colorlinks=true,citecolor=blue,linkcolor=magenta]{hyperref}
\usepackage[usenames]{color}

\begin{document}

\title{Teleportation-based realization of an optical quantum two-qubit entangling gate}

\author{Wei-Bo Gao}
\thanks{These authors contribute equally to this work}
\affiliation{Hefei National Laboratory for Physical Sciences at Microscale and Department of Modern Physics, University of Science and Technology of China, Hefei, Anhui 230026, China}
\author{Alexander M. Goebel}
\thanks{These authors contribute equally to this work}
\affiliation{Physikalisches Institut, Ruprecht-Karls-Universit\"{a}t Heidelberg, Philosophenweg 12, 69120 Heidelberg, Germany}
\author{Chao-Yang Lu}
\thanks{These authors contribute equally to this work}
\affiliation{Hefei National Laboratory for Physical Sciences at Microscale and Department of Modern Physics, University of Science and Technology of China, Hefei, Anhui 230026, China}
\author{Han-Ning Dai}
\affiliation{Hefei National Laboratory for Physical Sciences at Microscale and Department of Modern Physics, University of Science and Technology of China, Hefei, Anhui 230026, China}
\author{Claudia Wagenknecht}
\affiliation{Physikalisches Institut, Ruprecht-Karls-Universit\"{a}t Heidelberg, Philosophenweg 12, 69120 Heidelberg, Germany}
\author{Qiang Zhang}
\affiliation{Hefei National Laboratory for Physical Sciences at Microscale and Department of Modern Physics, University of Science and Technology of China, Hefei, Anhui 230026, China}
\author{Bo Zhao}
\affiliation{Hefei National Laboratory for Physical Sciences at Microscale and Department of Modern Physics, University of Science and Technology of China, Hefei, Anhui 230026, China}
\author{Cheng-Zhi Peng}
\affiliation{Hefei National Laboratory for Physical Sciences at Microscale and Department of Modern Physics, University of Science and Technology of China, Hefei, Anhui 230026, China}
\author{Zeng-Bing Chen}
\affiliation{Hefei National Laboratory for Physical Sciences at Microscale and Department of Modern Physics, University of Science and Technology of China, Hefei, Anhui 230026, China}
\author{Yu-Ao Chen}
\thanks{Current address: Fakult\"at f\"ur Physik, Ludwig-Maximilian-Universit\"at, Schellingstrasse 4, 80798 M\"unchen, Germany}
\affiliation{Hefei National Laboratory for Physical Sciences at Microscale and Department of Modern Physics, University of Science and Technology of China, Hefei, Anhui 230026, China}
\author{Jian-Wei Pan}
\affiliation{Hefei National Laboratory for Physical Sciences at Microscale and Department of Modern Physics, University of Science and Technology
of China, Hefei, Anhui 230026, China}
\affiliation{Physikalisches Institut, Ruprecht-Karls-Universit\"{a}t Heidelberg, Philosophenweg 12, 69120 Heidelberg, Germany}

\date{\today}

\begin{abstract}
In recent years, there has been heightened interest in quantum
teleportation, which allows for the transfer of unknown quantum
states over arbitrary distances. Quantum teleportation not only
serves as an essential ingredient in long-distance quantum communication,
but also provides
enabling technologies for practical quantum computation. Of
particular interest is the scheme proposed by Gottesman and Chuang
[Nature \textbf{402}, 390 (1999)], showing that quantum gates can be
implemented by teleporting qubits with the help of some special
entangled states. Therefore, the construction of a quantum computer
can be simply based on some multi-particle entangled states, Bell state
measurements and single-qubit operations. The feasibility of this scheme
relaxes experimental
constraints on realizing universal quantum computation. Using
two different methods we demonstrate the smallest non-trivial module
in such a scheme---a teleportation-based quantum entangling gate for
two different photonic qubits.
One uses a high-fidelity six-photon
interferometer to realize controlled-NOT gates and the other uses four-photon
hyper-entanglement to realize controlled-Phase gates. The results
clearly demonstrate the working principles and the entangling
capability of the gates. Our experiment represents an important step
towards the realization of practical quantum computers and could
lead to many further applications in linear optics quantum
information processing.
\end{abstract}

\maketitle

In 2001, Knill, Laflamme and Milburn (KLM) showed that scalable and
efficient quantum computation is possible by using linear optical
elements, ancilla photons and post-selection \cite{KLM}. The KLM
scheme is based on three principles. First, non-deterministic
quantum computation is possible with linear optics. Second,
universal quantum gates with the probability approaching one can be
implemented based on teleportation \cite{gottesman99}, a process in
which a qubit in an unknown state can be transferred to another
qubit \cite{bennett93,bouwmeester}. Third, the demanding resources
can be reduced by quantum coding. The first principle has been
demonstrated in many experiments. For example, various approaches
for realizing photonic controlled-NOT (C-NOT) gates have been
reported \cite{brien03,gasparoni04,fio04,Langford,Kiesel,Okamoto,
Vallone}. Recently, a three-qubit Toffoli gate has also been carried
out in a photonic architecture \cite{whiteto}. Additionally,
there have been many efforts aimed at reducing the resource
requirements of the KLM protocol \cite{yoran03,
Nielsen04,browne05,gilchrist05}. Nevertheless, the
teleportation-based two-qubit entangling gate, which plays an
important role in the second principle of the KLM scheme, still
remains an experimental challenge.

Quantum teleportation is useful for quantum communication
\cite{repeater,gisin02} since it allows us to use entangled states as perfect
quantum channels. The novel scheme proposed by
Gottesman and Chuang (GC) in 1999 \cite{gottesman99} opens the way
for promising applications in realizing quantum computation (QC)
\cite{KLM,gottesman99,yoran03,Nielsen04,Nielsen03}. In the GC
scheme, qubits are teleported through special gates by simply using
multi-particle off-line entangled states, Bell state measurements
(BSM) and single-qubit operations. It can be extended to
implement universal measurement-based quantum computation. For
example, in Refs. \cite{yoran03,Nielsen03}, joint two-qubit
measurements have been used to implement a teleportation-based model
of quantum computation. It has also been shown that one-way quantum
computation based on cluster states \cite{Robert01} is equivalent
with the teleportation-based approaches \cite{Nielsen05,
Verstraete04}. In addition, the GC scheme can be used to implement a
nearly deterministic quantum gate. It teleports the qubits through a
non-deterministic gate that has already been realized. Using more and
more qubits, an entangling gate with a probability of success
approaching one can be implemented \cite{KLM}.

To implement the fundamental building block of the GC scheme,
teleportation-based C-NOT gate or controlled-Phase (C-Phase) gate,
one has to use at least six qubits. All logic operations needed for
quantum computation can be performed using single qubit operations
in combination with a  C-NOT gate or C-Phase gate \cite{Barenco95b}. In our
experiment, we first realize a teleportation-based C-NOT gate with
six photons. We measure the fidelities for the truth table of
the gate and an entangled output state. Next, we implement a
C-Phase gate by using a four-photon hyper-entangled state
\cite{Kwiat}. The fidelity of the gate is estimated. Moreover,
we show that quantum parallelism is achieved in our C-phase
gate, thus proving that the gate can not be reproduced by local
operations and classical communications \cite{hofmann02}. Our
experiment represents a non-trivial proof-of-principle
implementation of the teleportation protocol introduced by Gottesman
and Chuang.

\section{Theoretical Schemes}

\begin{figure}
\begin{center}
\includegraphics[width=8cm]{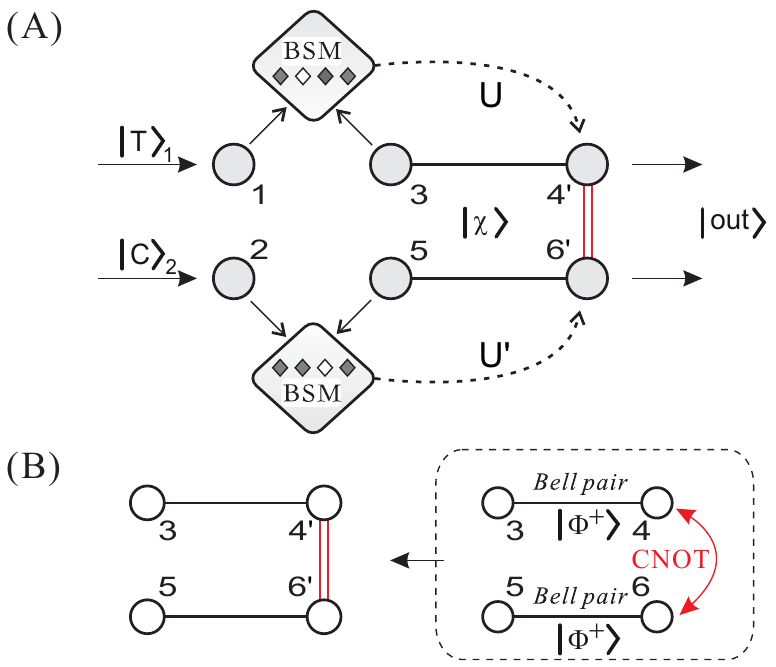}
\caption{Quantum circuit for teleporting two qubits through a C-NOT
gate and a C-Phase Gate. (\emph{A}) The input consisting of the
target qubit $|T\rangle_{1}$ and control qubit $|C\rangle_{2}$ can
be arbitrarily chosen. Bell State Measurements (BSMs) are performed
between the input states and the left qubits of the special
entangled state $|\chi\rangle$. Depending on the outcome of the
BSMs, local unitary operations (U, U$^{\prime}$) are applied on the remaining
qubits of $|\chi\rangle$, which then form the output $|out\rangle =
U^{C-NOT} |T\rangle_{1}|C\rangle_{2}$ or $|out\rangle =
U^{C-Phase} |T\rangle_{1}|C\rangle_{2}$. (\emph{B}) The
special entangled state $|\chi\rangle$ can be constructed by
performing a C-NOT gate on two Bell pairs, with $|\Phi^{+}\rangle =
\frac{1}{\sqrt{2}} \left(|H\rangle|H\rangle +
|V\rangle|V\rangle\right)$. See Appendix for details.}
\end{center}
\end{figure}

A key element in the scheme of Gottesman and Chuang is to implement
the C-NOT gate, which acts on a control and a target qubit. Here the
logic table of the C-NOT operation ($U^{C-NOT}$) is given by
$|H\rangle_{1}|H\rangle_{2} \rightarrow |H\rangle_{1}|H\rangle_{2}$,
$|H\rangle_{1}|V\rangle_{2} \rightarrow |V\rangle_{1}|V\rangle_{2}$,
$|V\rangle_{1}|H\rangle_{2} \rightarrow |V\rangle_{1}|H\rangle_{2}$
and $|V\rangle_{1}|V\rangle_{2} \rightarrow
|H\rangle_{1}|V\rangle_{2}$, where we have encoded qubits on the
polarization degree of freedom of photons. A schematic diagram of
the procedure is shown in Fig.~1\emph{A}. First, we prepare
beforehand an entangled four-qubit state $|\chi\rangle$. Next, by
using quantum teleportation, we transfer the data of the two input
qubits $|T\rangle_1$ (target) and $|C\rangle_2$ (control) onto
$\vert\chi\rangle$. Specifically, this is done by projecting the target
(control) qubit and one of the outer qubits of $|\chi\rangle$ onto a
joint two-particle ``Bell state''. To finish the procedure, we apply
single qubit (Pauli) operations to the output qubits
depending on the outcomes of the BSMs. Now the output is in the desired
state given by
\begin{equation}
|out\rangle= U^{C-NOT} |T\rangle_{1}|C\rangle_{2}.\label{out}
\end{equation}
This can be better understood by a closer look at the special
entangled state $|\chi\rangle$. It is a four-particle cluster state
\cite{Robert2} of the form
\begin{eqnarray}
|\chi\rangle = \frac{1}{2} [ (|H\rangle_{3}|H\rangle_{4^{\prime}} +
|V\rangle_{3}|V\rangle_{4^{\prime}}) |H\rangle_{5}|H\rangle_{6^{\prime}} \nonumber\\
+ (|H\rangle_{3}|V\rangle_{4^{\prime}} +
|V\rangle_{3}|H\rangle_{4^{\prime}})
|V\rangle_{5}|V\rangle_{6^{\prime}}],
\end{eqnarray}
which can be created simply by performing a C-NOT operation on two
EPR pairs $|\Phi^{+}\rangle = \frac{1}{\sqrt{2}}
\left(|H\rangle|H\rangle + |V\rangle|V\rangle\right)$ (see
Fig.~1\emph{B}). Note that application of this C-NOT
operation onto the two EPR pairs prior to teleportation is the
reason that the input states have undergone a CNOT gate after
teleportation. This is the essential difference between our scheme
and standard teleportation. A detailed discussion of the scheme is
shown in the Appendix.


When the off-line entangled resource is prepared in a different
state, we can teleport the input qubits through a different
entangling gate. For example, we prepare the off-line state as
\begin{eqnarray}\label{1} \left| {\chi^{\prime}} \right\rangle
&=& \frac{1} {2}[ (\left| H \right\rangle_3 \left| H \right\rangle_{4^{\prime}}  +
\left| V \right\rangle_3 \left| V \right\rangle_{4^{\prime}}
)\left| H \right\rangle_{5} \left| H \right\rangle_{6^{\prime}} \notag\\
&& +  (\left| H \right\rangle_3 \left| H \right\rangle_{4^{\prime}}
- \left| V \right\rangle_3 \left| V \right\rangle_{4^{\prime}}
)\left| V \right\rangle_{5} \left| V \right\rangle_{6^{\prime}}],
\end{eqnarray}
which results from performing a C-Phase
gate between two EPR pairs $|\Phi^{+}\rangle$. Here the logic table
of the C-Phase operation is given by $|H\rangle|H\rangle \rightarrow
|H\rangle|H\rangle$, $|H\rangle|V\rangle \rightarrow
|H\rangle|V\rangle$, $|V\rangle|H\rangle \rightarrow
|V\rangle|H\rangle$ and $|V\rangle|V\rangle \rightarrow
-|V\rangle|V\rangle$. After the creation of the entangled state
$\left| {\chi^{\prime}} \right\rangle$, we implement BSMs on qubit
$|T\rangle_1$ with qubit $3$, and qubit $|C\rangle_2$ with qubit
$5$. Based on the results of the BSMs, 16 possible Pauli corrections
(See \emph{Appendix}) are applied on qubits $4^{\prime}$ and
$6^{\prime}$. This allows us to teleport the two qubits through a
C-Phase gate. Note that the two types of entangling gates are
equivalent up to single qubit unitary transformations. For example,
by applying two Hadamard gates on one of the input qubits and the
corresponding output qubit after teleportation, the C-Phase gate can
be converted to a C-NOT gate.

\section{The experimental demonstration with six photons}

\begin{figure}
\includegraphics[width=3in]{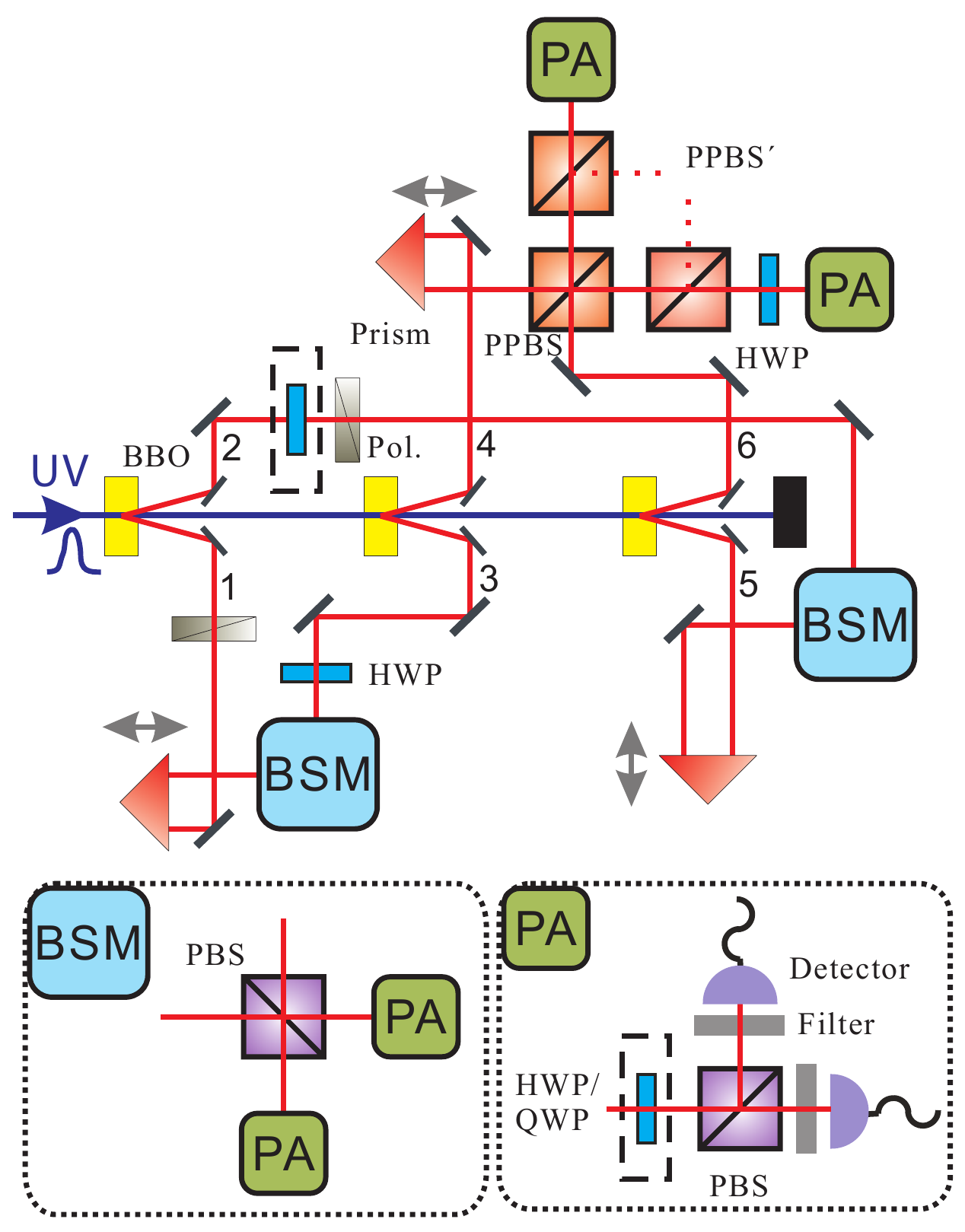}\\
\caption{A schematic diagram of the experimental setup. We
frequency double a mode-locked Ti:sapphire laser system to
create a high-intensity pulsed ultraviolet (UV) laser beam
at a central wavelength of 390 nm, a pulse duration of 180 fs, and a
repetition rate of 76 MHz. The UV beam successively passes through three
$\beta$-barium borate (BBO) crystals to generate three polarization
entangled photon pairs via type-II spontaneous parametric
down-conversion \cite{Kwiat95}. At the first BBO the UV generates a
photon pair in modes 1 and 2 (i.e. the input consisting of the
target and control qubit). After the crystal, the UV is refocused
onto the second BBO to produce another entangled photon pair in
modes 3 and 4 and correspondingly for modes 5 and 6. Photons 4 and 6
are then overlapped at a PPBS and together with photons 3 and 5
constitute the cluster state. Two PPBS' are used for state
normalization. The prisms are mounted on step motors and are used to
compensate the time delay for the interference at the PPBS and the
BSMs. A BSM is performed by overlapping two incoming photons on a
PBS and two subsequent polarization analyses (PA). A PA projects the
photon onto an unambiguous polarization depending on the basis
determined by a half or quarter wave plate (HWP or QWP).
The photons are detected by silicon avalanched single-photon
detectors. Coincidences are recorded with a coincidence unit clocked
by the infrared laser pulses. Polarizers (Pol.) are polarizers used
to prepare the input state and narrow band filters (Filter) with
$\Delta_{FWHM}=3.2$ nm are used to obtain a better spectral interference.
}\label{fig2}
\end{figure}

\subsection{The creation of the four-photon state $|\chi\rangle$}

The C-NOT gate is implemented by using six photons.  A schematic
diagram of the  experimental setup is shown in Fig.~2. All three
photon pairs are originally prepared in the Bell-state
$|\Phi^{+}\rangle = \frac{1}{\sqrt{2}}(|H\rangle|H\rangle +
|V\rangle|V\rangle)$. We observe on average $7\times10^4$ photon
pairs per second from each (EPR) source with a visibility of 87.5
\%. With this high-intensity entangled photon source we obtain in
total 3.5 six-photon events per minute. This is less than half the
count rate of our previous six-photon experiments
\cite{Qiang06Teleportation,Chaoyang07}. Since the new scheme is more
complex and involves more interferences, the fidelity requirements
are more stringent. Thus, we reduce the pump power from 1.0
W to 0.8 W in order to reduce noise contributions from
the emission of two pairs of down-converted photons by a single
source (double-pair-emission).

Using wave plates and polarizers, we prepare photon pair
1\&2 in the desired two-qubit input state $|\Psi\rangle_{12}$.
Photon pairs 3\&4 and 5\&6, which are both in the state
$|\Phi^{+}\rangle$, are used as resources to construct the special
entangled state $|\chi\rangle$. In the experiment, we use a two
photon C-NOT gate to produce the desired cluster state
\cite{Hofmann, Langford, Kiesel, Okamoto}. As shown in
Fig.~2, photons 4 and 6 are interfered on partially polarizing
beam splitters (PPBS), i.e. the transmission for the horizontal
(vertical) polarization is $T_H=1$ ($T_V=1/3$). In order to balance
the transmission for all input polarizations, PPBS' with reversed
transmission conditions ($T_H = 1/3$, $T_V = 1$) are placed in each
output of the overlapping PPBS. Altogether, the probability of
having one photon in each desired output, and thus of having
successfully created the cluster state, is $1/9$. Half wave plates
(HWPs) in arms 3 and 4 are used to transform the cluster state to
the desired state by local unitary operations.

To achieve good spatial and temporal overlap, the photons are
spectrally filtered with very steep edge narrow-band filters
($\Delta\lambda_{FWHW} = 3.2$ nm) and detected
by fibre-coupled single-photon detectors. At the same time, by
shortening the distance between the BBO and the fiber
coupler, and carefully refocusing the UV pulse with appropriate
lenses, we are able to obtain an overall efficiency of about $15\%$
(including the coupling and detection efficiency). The experimental
count rate for the four-qubit cluster state $|\chi\rangle$ is
$~7/s$, which is two orders of magnitude larger than in a recent
experiment \cite{Kiesel05}. Using the same method as
ref. \cite{Kiesel05} , we measure the state fidelity to be
$0.694 \pm0.003$, which is slightly lower than that in \cite{Kiesel05}.
The imperfect preparation of the desired cluster state is mainly
limited by high-order emissions of entangled photons, the imperfect
interference on PPBS and as well as the quality of
PPBS, whose transmission ratio for different polarization is
measured to be about $T_H=0.95$ and $T_V=0.3$ for one input and
$T_H=0.96$ and $T_V=0.35$ for the other input.

\subsection{Teleporting two qubits through a C-NOT gate}

\begin{figure}
\includegraphics[width=3.3in]{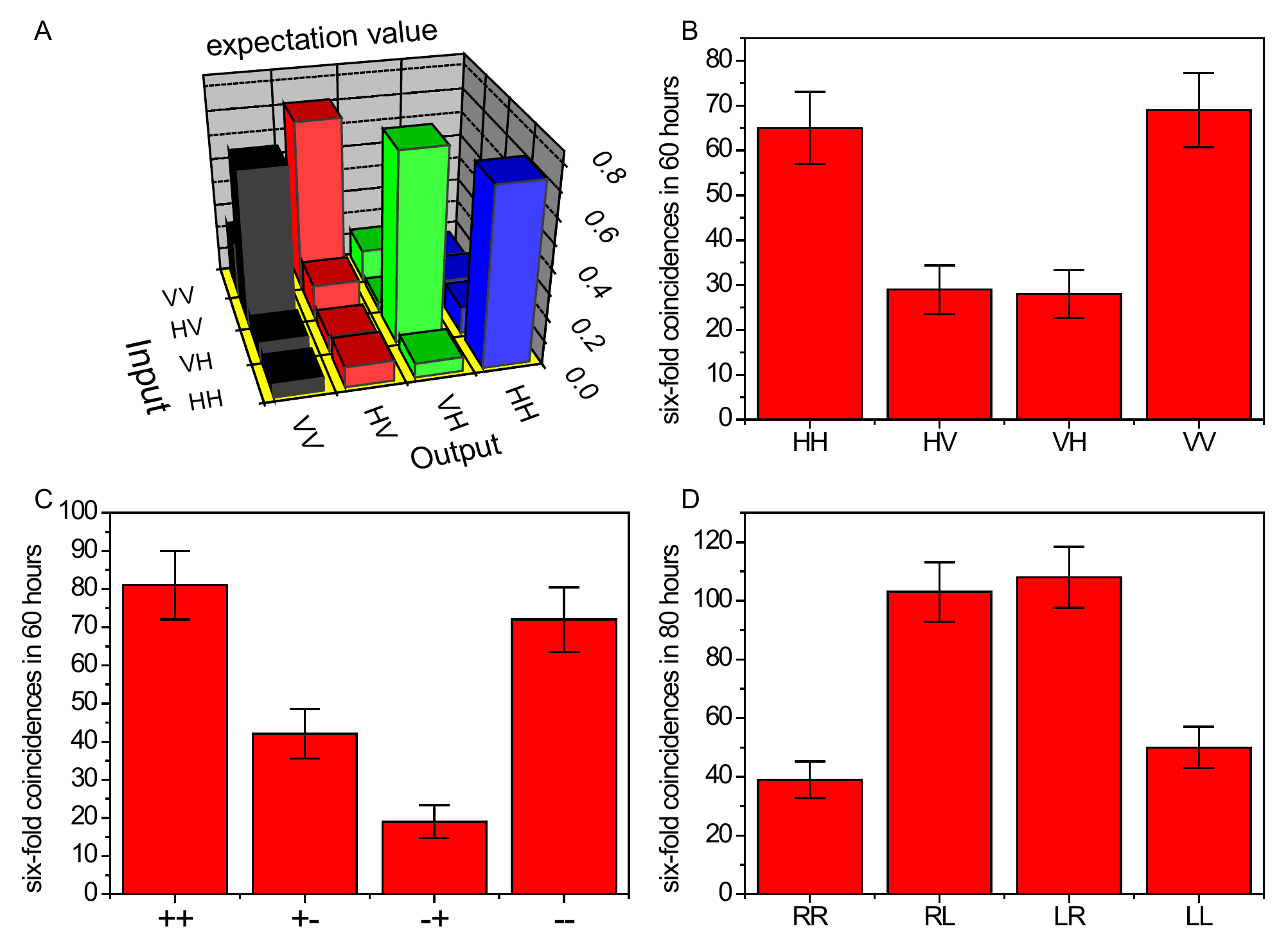}\\
\caption{Experimental results for teleportation-based C-NOT gate.
(\emph{A}) Experimental results for truth table of the C-NOT gate.
The first qubit is the target and the second is the control qubit.
In the data, we considered the corresponding unitary transformation
depending on the type of coincidence at the BSM
($|+\rangle|+\rangle$, $|+\rangle|-\rangle$, $|-\rangle|+\rangle$,
$|-\rangle|-\rangle$). The average fidelity for the truth table is
$0.72 \pm 0.05$. (\emph{B}) Experimental results for fidelity
measurements of entangled output states. Basis $|H\rangle/|V\rangle$
is used for the measurements of $\langle \hat{\sigma}_z
\hat{\sigma}_z \rangle$; (\emph{C}) $|+\rangle/|-\rangle$ for
$\langle \hat{\sigma}_x \hat{\sigma}_x \rangle$; (\emph{D})
$|L\rangle/|R\rangle = \frac{1}{\sqrt{2}} \left(|H\rangle \pm
i|V\rangle \right)$ for $\langle \hat{\sigma}_y \hat{\sigma}_y
\rangle$. The measured expectation values are: (\emph{B}) $0.403 \pm
0.066$ (\emph{C}) $0.462 \pm 0.057$ and (\emph{D}) $-0.434 \pm
0.062$. All errors are statistical and correspond to $ \pm
1$ standard deviations.}\label{fig3}
\end{figure}

Teleporting the input data of $|\psi\rangle_{12}$ to $|\chi\rangle$
requires joint BSMs on photons 1\&3 and photons 2\&5. To demonstrate
the working principle of the teleportation-based C-NOT gate, it is
sufficient to identify one of the four Bell states in both BSMs
\cite{Qiang06Teleportation,Goebel08}. However, in the experiment we
choose to analyse the two Bell states $|\Phi^{+}\rangle$ and
$\vert\Phi^-\rangle$ to increase the efficiency - the fraction of
success - by a factor of 4. This is achieved by interfering photons
1\&3 and photons 2\&5 on a polarizing beam splitter (PBS) and
performing a polarization analysis (PA) on the two outputs
\cite{Pan98GHZa}. With the help of a HWP, a PBS and fibre-coupled
single photon detectors, we are able to project the input photons of
the BSM onto $|\Phi^{+}\rangle$ upon the detection of a
$|+\rangle|+\rangle$ or $|-\rangle|-\rangle$ coincidence, and onto
$|\Phi^{-}\rangle$ upon the detection of a $|+\rangle|-\rangle$ or
$|-\rangle|+\rangle$ coincidence (where $|\pm\rangle=(|H\rangle \pm
|V\rangle)/\sqrt{2}$). The increase in success efficiency in comparison with
Ref. \cite{Qiang06Teleportation,Goebel08} allows us to reduce
the pump power in order to reduce noise contributions while
preserving the overall count rate.  Note that even with a 1/4 success
probability of the BSM, the in-principle demonstration of the protocol
will not be affected since the unsuccessful measurements can
be thought of as a photon loss error and will not affect the
fidelity of final output. Furthermore, when including the
nonlinearity of the detection process, it is, in principle, possible
to construct complete BSMs with increasing resources \cite{Kok07}.

\subsection{Experimental results} The projective BSMs between the data input photon 1 (2)
and photon 3 (5) of the cluster state leave the remaining photons of
the cluster state 4\&6 in the state $\vert out\rangle_{46}$ up to a
unitary transformation. This is the desired final state after
performing a C-NOT operation on photons 1\&2. To demonstrate that our
teleportation-based C-NOT gate protocol works for a general unknown
polarization state of photons 1\&2, we measure the truth table of
our gate. That is, we measure the output for all possible
combinations of the two-qubit input in the computational basis.
However, this is not sufficient to show the quantum characteristic
of a C-NOT gate. The remarkable feature of a C-NOT gate is its
ability to entangle two separable qubits. Thus, to fully
demonstrate the successful operation of our protocol, we
perform the entangling operation:
\begin{eqnarray}
&|H\rangle_T \otimes \frac{1}{\sqrt 2}(|H\rangle_C + |V\rangle_C)
\stackrel{C-NOT}{\longrightarrow} \nonumber\\ &\frac{1}{\sqrt
2}(|H\rangle_T|H\rangle_C + |V\rangle_C|V\rangle_C) =
|\Phi^+\rangle_{TC}
\end{eqnarray}
We quantify the quality of our output state by looking at the
fidelity as defined by $F=Tr(\hat\rho\vert out\rangle\langle
out\vert)$ where $\vert out \rangle$ is the theoretically desired
final state and $\hat\rho$ is the density matrix of the experimental
output state.

Here, we discuss the fidelity measurements for the truth table.
Conditional on detecting a fourfold coincidence of the
two BSMs, we analyze the output photons 4\&6 in the computational
$H/V$ basis. The measurement results are shown in Fig.~3\emph{A},
where the corresponding unitary transformation according to
different results of the BSMs have been considered. The experimental
integration time for each possible combination of the input photons
was about 50 hours, and we recorded about 120 desired two-qubit
events. In the experiment, we obtained an average fidelity of $
F_{z4'z6'}= 0.72 \pm 0.05$ for the output states of the truth table,
which is defined as
\begin{eqnarray}
 F_{z4'z6'}  &=& 1/4[P(HH|HH) + P(VH|VH)\notag\\
&&+ P(VV|HV) + P(HV|VV)]. \notag\\
 \end{eqnarray}
Here $P$ represents the probability of obtaining the corresponding
output state under the specified input state. For example,
$P(VV|HV)$ represents the probability of the case that the output
state is $|V\rangle_{4'}|V\rangle_{6'}$ when the input state is
$|H\rangle_1|V\rangle_2$.

Next, we demonstrate the entangling capability of the gate.  This
can be seen by a closer look at the fidelity:
\begin{eqnarray}\label{Fidelity}
F & = & Tr(\hat{\rho}|\Phi^{+}\rangle\langle\Phi^{+}|) \nonumber\\
{} & = & \frac{1}{4} Tr \left(\hat{\rho}(\hat{I} + \hat{\sigma}_{x}
\hat{\sigma}_{x} - \hat{\sigma}_{y} \hat{\sigma}_{y} +
\hat{\sigma}_{z} \hat{\sigma}_{z}) \right).
\end{eqnarray}
This implies that by measuring the expectation values
$\langle\hat{\sigma}_{x}\hat{\sigma}_{x}\rangle$,
$\langle\hat{\sigma}_{y}\hat{\sigma}_{y}\rangle$,
$\langle\hat{\sigma}_{z}\hat{\sigma}_{z}\rangle$, we can directly
obtain the fidelity of the entangled output state. The experimental
results for the correlated local measurement settings are
illustrated in Fig.~3. The integration time for the first two
settings was about 60 hours and for the third setting about 80
hours. Using Eq.~\ref{Fidelity}, we find a fidelity
of $0.575 \pm 0.027$, which is above 0.50 and
thus proves genuine two-photon entanglement in the output states
\cite{Guehne02}.

All experimental results are calculated directly from the original
data and no noise contributions have been subtracted. The
imperfect fidelity is due to several reasons. First, the
imperfect preparation of the cluster state is the main limitation
for the non-ideal C-NOT gate. Second, the large pump power
double-pair-emission contributes significantly to the noise, which can be
seen from the reduction of teleportation fidelity with or without
the third pair. Furthermore, the interference visibility is limited
since the complex phase compensations drift over the long
measurement times. Imperfect input states also reduce the quality of
our output states. Note that we achieve a better fidelity for the
truth table than for the entangling case. This is because in the
latter case, the fidelity depends on the interference visibility at
the PBS of the BSM. All given errors are of statistical nature and
correspond to $\pm 1$ standard deviations.

\section{Demonstration of the C-Phase gate with hyperentanglement}

\begin{figure}
\includegraphics[width=3.3in]{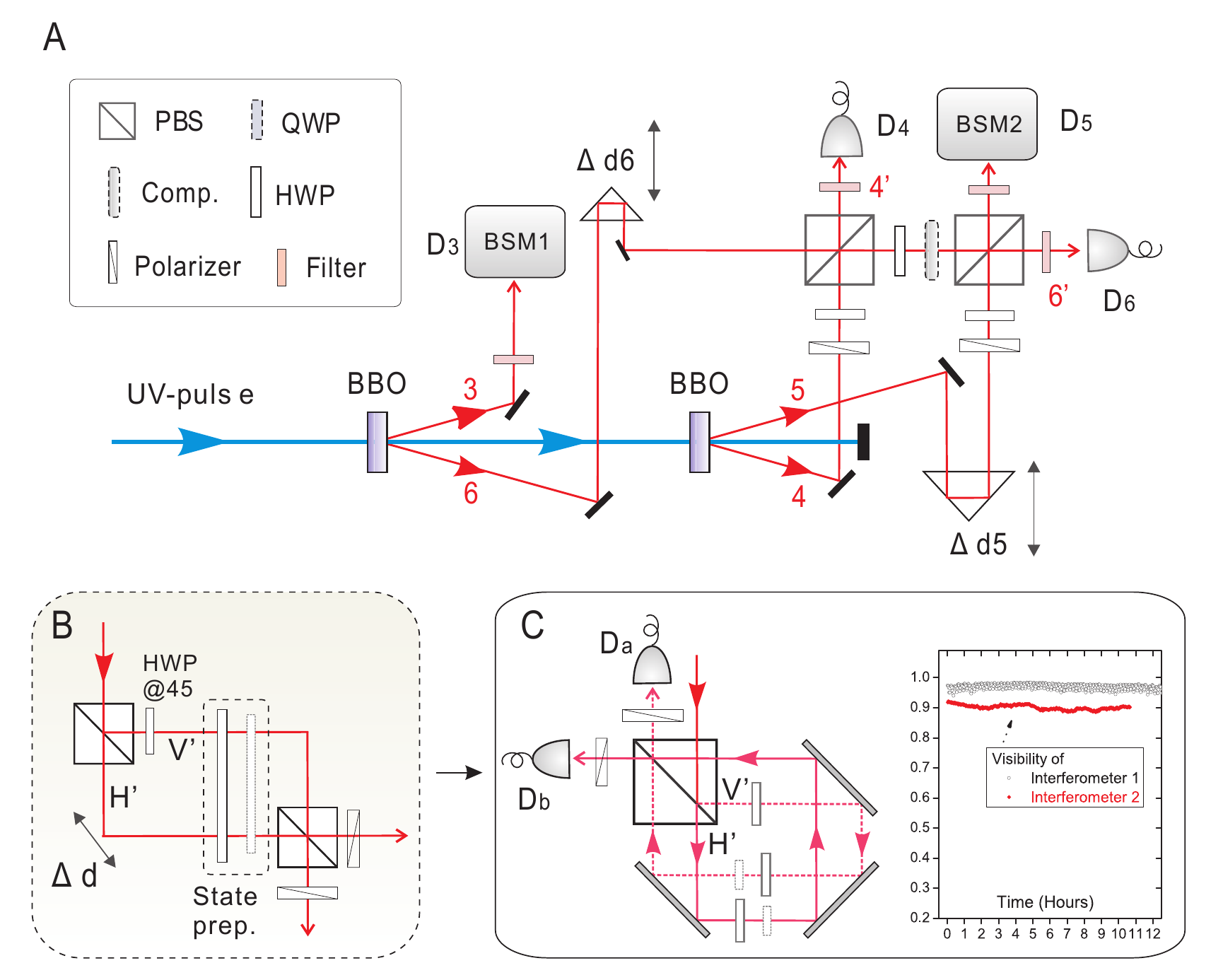}\\
\caption{Schematic of the experimental setup. (\emph{A})
Femtosecond UV pulses pass through two BBO  crystals to create two pairs of entangled photons.
Two polarizers are inserted in the arms of $3$ and $4$ to prepare single photons in
$|+\rangle=(1/\sqrt{2})(|H\rangle+|V\rangle)$.
(\emph{B}) Photons $3$ and $5$ are sent through  Mach-Zehnder-type
interferometers to perform the spatial-polarization bell-state
measurement (BSM). Polarization and spatial qubit transformation happens at the first
PBS, and BSM happens at the second PBS. (\emph{C}) In the experiment, we
use an ultra-stable Sagnac configuration interferometer to satisfy
the desired high stability.}\label{33}
\end{figure}

With the help of hyper-entanglement, we are able to tackle the
problem of low counting rates in the six-photon experiment
\cite{Tokunaga}. More importantly, it is proved that the  GC scheme
with hyperentanglement can be extended to implement universal
quantum computation based on the so-called ``linked-state''
\cite{yoran03}. The ``linked-state'' consists of chains of
photons. Every single logical qubit corresponds to a chain, where
the spatial degree of freedom of each photon is maximally entangled
with the polarization of the next photon. The chains are linked
according to the circuit which one wishes to process. Once this
state has been successfully prepared, the computation can be
realized deterministically by a sequence of teleportation steps and
complete single-photon spatial-polarization BSMs \cite{yoran03,Mor}.

\subsection{The creation of the hyper-entangled four-qubit state $\left| {\chi'} \right\rangle$}

\begin{figure}
\includegraphics[width=3.3in]{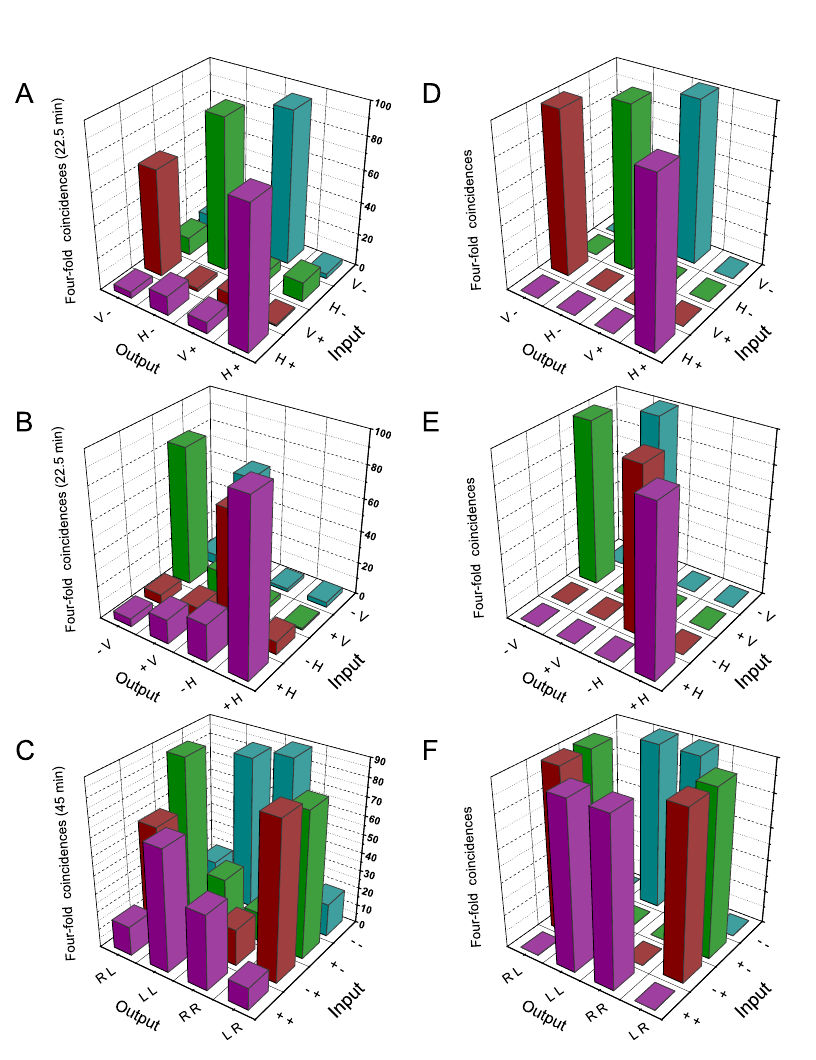}
\caption{Experimental evaluation of the quality of the C-phase gate.
Data for $F_{z4'x6'}$ and $F_{x4'z6'}$ are measured for
22.5 minutes respectively, and data for $F_{x4'x6'}$ are measured for
45 minutes. ($A$) Experimental values for measurements of $F_{z4'x6'}$.
($B$) Experimental values for measurements of $F_{x4'z6'}$.
($C$) Experimental values for measurements of $F_{x4'x6'}$.
($D$) Theoretical values for $F_{z4'x6'}$.
($E$) Theoretical values for $F_{x4'z6'}$.
($F$) Theoretical values for $F_{x4'x6'}$. }\label{1}
\end{figure}

In the experiment, we use $\left| {\chi'} \right\rangle$ (Eq. (3))
to teleport the input qubits through a C-Phase gate. In $\left|
{\chi'} \right\rangle$, qubits 3, 5 are encoded on the spatial modes of
photons and qubits 4$^{\prime}$, 6$^{\prime}$ are encoded on the
polarization degree of freedom of photons (See Fig. 1\emph{A}). A
schematic of the experimental setup is shown in Fig. 4. A pulsed
ultraviolet laser beam passes through two BBO crystals to create two
pairs of entangled photons. The first pair is prepared in the state
$|\Phi^{+}\rangle_{36}$. The second pair is disentangled with
polarizers and initialized in the state $\left| +
\right\rangle_{4}\left| + \right\rangle_{5}$. Then photons 4, 6 and
5, 6 are overlapped on two polarizing beam splitters (PBS) to
prepare a four-photon entangled state \cite{yamo}:
\begin{eqnarray}\label{1}
\left| {\lambda} \right\rangle  &=& \frac{1} {2}[\left| H
\right\rangle_3 \left| H \right\rangle_{4^\prime} (\left| H \right\rangle_{5}
\left| H\right\rangle_{6^{\prime}} + \left| V \right\rangle_{5} \left| V
\right\rangle_{6^{\prime}}
)\notag\\
&& + \left| V \right\rangle_3 \left| V \right\rangle_{4^\prime} (\left| H
\right\rangle_{5} \left| H \right\rangle_{6^{\prime}}  - \left| V \right\rangle_{5}
\left| V \right\rangle_{6^{\prime}} )].
\end{eqnarray}
We find the fidelity $ F
=Tr(\left|\lambda\right\rangle\left\langle\lambda\right|\rho_{exp})$
of the prepared state to be $0.71\pm 0.01 $, which is above 0.5 by 21 standard deviations and
thus proves the genuine four-qubit entanglement in the state
\cite{fidelity}. Based on the state $\left| {\lambda}
\right\rangle$,  we place a PBS in each output of photons 3 \& 5.
Since the PBS transmits $H$ and reflects $V$ polarization, the
$H$-polarized photon will go to one path and $V$-polarized photon
will go to the other path. If we denote the levels of spatial qubits
as $\left|H^{\prime}\right\rangle$ for the first path and
$\left|V^{\prime}\right\rangle$ for the second path, $\left|
{\lambda} \right\rangle$ will be converted to:
\begin{eqnarray}\label{1}
\left| \widetilde{\chi} \right\rangle  &=& \frac{1} {2}[\left|
H^{\prime} \right\rangle_{3} \left| H \right\rangle_{4^\prime}
(\left| H^{\prime}\right\rangle_5 \left| H \right\rangle_{6^\prime}
+ \left| V^{\prime} \right\rangle_5 \left| V
\right\rangle_{6^\prime}
)\notag\\
&& + \left| V^{\prime} \right\rangle_3 \left| V
\right\rangle_{4^\prime} (\left| H^{\prime} \right\rangle_5 \left| H
\right\rangle_{6^\prime}   - \left| V^{\prime} \right\rangle_5
\left| V \right\rangle_{6^\prime} )],
\end{eqnarray}
which is equivalent to $\left| {\chi'} \right\rangle$, except for
that qubits 3 and 5 are defined on the spatial degrees of
freedom of photons.

\subsection{Teleporting two qubits through a C-Phase gate}

\begin{figure}
\includegraphics[width=3.3in]{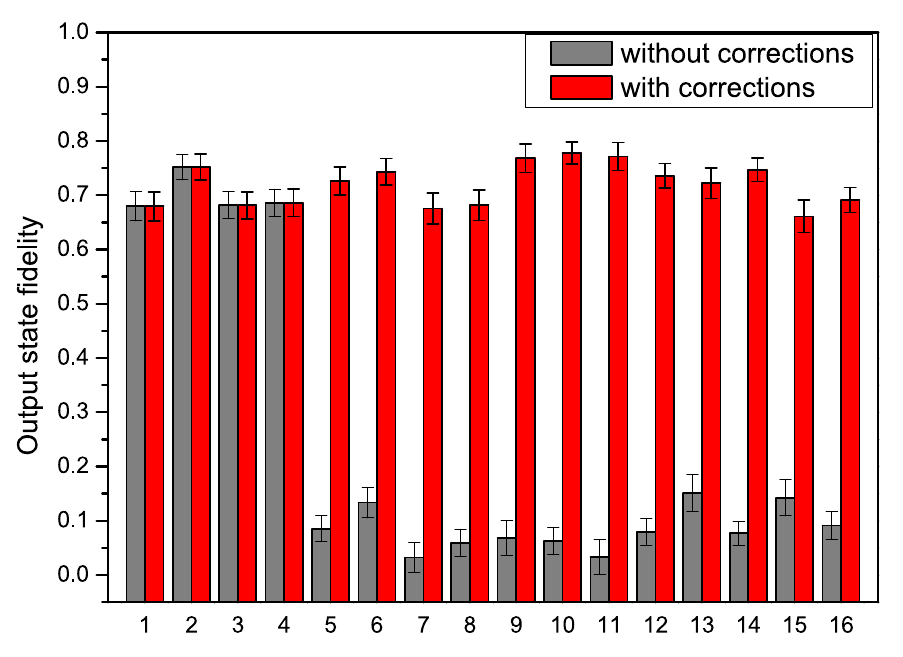}\\
\caption{The fidelity with the expected state before and after the
correction operations. The input control and target qubit are
both in the state $ \left| + \right\rangle$, so the output state is expected to be
$\frac{1}{{\sqrt 2 }}(\left| {H+} \right\rangle  + \left| {V-}
\right\rangle )$. The fidelity is much higher after correction operations.
The 16 cases correspond to the 16 different outputs of the two BSMs (see TABLE I
in \emph{Appendix})}\label{1}
\end{figure}

We now discuss the preparation of input qubits
and the implementation of BSMs. In the experiment, the polarization
mode of photon 3 is used as the input target qubit  $ \left| T
\right\rangle_1$ and the polarization mode of photon 5 is used as
the input control qubit $ \left| C \right\rangle_2$. As shown in
Fig. 4\emph{B}, by placing HWPs oriented at $45^{\circ}$ with
respect to the horizontal direction in the spatial mode
$|V'\rangle_3$ and $|V'\rangle_5$, the $|V\rangle$ component of
photons 3 and 5 will be converted to $|H\rangle$. In this way, the
qubits $ \left| T \right\rangle_1$ and $ \left| C \right\rangle_2$
to be teleported are both prepared in the initial state
$\left|H\right\rangle$. Then, by using a combination of HWPs and
QWPs, we can prepare arbitrary input states $
\left| T \right\rangle_1$ and $ \left| C \right\rangle_2 $. The
required complete spatial-polarization BSMs are realized by two
single photon interferometers (see Fig. 4\emph{B}). Here Bell states
are denoted as
\begin{eqnarray}\label{2}
\left|\Phi^ \pm\right\rangle _{i} = \frac{1}{{\sqrt 2 }}(\left| H
\right\rangle _i \left| H^{\prime} \right\rangle _i  \pm \left| V
\right\rangle _i \left| V^{\prime}
\right\rangle _i ),\notag\\
 \left|\Psi^ \pm
\right\rangle_i   = \frac{1}{{\sqrt 2 }}(\left| H \right\rangle _i
\left| V^{\prime} \right\rangle _i  \pm \left| V \right\rangle _i
\left| H^{\prime}\right\rangle _i ),
\end{eqnarray}
where $i= 3, 5$. By matching the two spatial modes in the Bell
states at a PBS, $\left|\Phi^ \pm\right\rangle _{i} $ will appear as
$\left|\pm\right \rangle_i$ in one output port of the PBS, while
$\left|\Psi^ \pm \right\rangle_i $ will appear as $\left|\pm\right
\rangle_i$ in the other output port. Experimentally, we discriminate
these situations to implement a complete BSM and make the
corresponding corrections according to different measurement results
(see TABLE I in \emph{Appendix}).

Since single photon interferometers are required to implement
BSMs, in the experiment we utilize an ultra-stable Sagnac
configuration interferometer \cite{Nagata,Almeida,Gao} to satisfy
the desired high stability. As depicted in Fig. 4\emph{C}, the $H$
component of the input qubit is transmitted and propagates through
the interferometer in the counterclockwise direction, while the $V$
component is reflected and propagates through the interferometer in
the clockwise direction. If the setup is adjusted well, the
interference will occur when the two spatial modes match
at the same PBS. Experimentally, the interferometers can be
ultra-stable for about ten hours \cite{Gao}.

\subsection{Experimental results} To evaluate the performance of the C-phase gate, we obtain
the upper and lower bounds of the quantum process fidelity and
entangling capability with a recent method \cite{hofmann01}. Let us
define the fidelities  as
\begin{eqnarray}
 F_{z4'x6'}  &=& 1/4[P(H+|H+) + P(H-|H-)\notag\\
&&+ P(V-|V+) + P(V+|V-)], \notag\\
 F_{x4'z6'}  &=& 1/4[P( +  H | +  H ) + P( -  V | +  V ) \notag\\
&&+ P( -  H | -  H ) + P( +  V | -  V )],
 \end{eqnarray}
where each $P$ has the same definition as in Eq. (5) . When the
results of Bell state measurements are $\left|\Phi^{+}\right\rangle
_3 $ and $\left|\Phi ^{+}\right\rangle_5$, the experimental results
to calculate $F_{z4'x6'}$ and $F_{x4'z6'}$ are depicted in Fig.
5\emph{A} and 5\emph{B}. In our experiment, $F_{z4'x6'}$ and
$F_{x4'z6'}$ result in $0.79 \pm 0.02$ and $0.82 \pm 0.02$,
respectively. The upper and lower bounds of the gate fidelity can be
obtained from these two fidelities as follows \cite{hofmann01}:
\begin{equation}\label{33}
    (F_{z4'x6'}  + F_{x4'z6'}  - 1) \le F_{process} \le \min (F_{z4'x6'} ,F_{x4'z6'} ).
\end{equation}
By substituting the experimental results into the above inequality,
we obtain the result that the gate fidelity lies between $0.61 \pm 0.03$
and  $0.79 \pm 0.02$. Since the fidelity of entanglement
generation is at least equal to the process fidelity, the lower
bound above also defines the lower bound of the gate's
\emph{entanglement capability}. In terms of the concurrence $C$ that
the gate can generate from product state inputs, the minimal
entanglement capability  is depicted by \cite{hofmann01}:
\begin{equation}\label{33}
   C \geq 2F_{process}-1 \geq 2(F_{z4'x6'}  +
F_{x4'z6'})-3.
\end{equation}
We obtain the result that the lower bound of the concurrence is
$0.22 \pm 0.06$, which is larger than zero and thus sufficient to
confirm the entanglement capability of our gate. The imperfection of
our gate is mainly due to undesired $H/V$ components caused by
high-order photon emissions and partial distinguishability of
independent photons interfered on the PBSs.

Furthermore, we demonstrate that quantum parallelism  has been
achieved in our C-phase gate, thus proving that the gate can not be
reproduced by local operations and classical communications
\cite{hofmann02}. As discussed in ref. \cite{hofmann02}, quantum
parallelism is achieved if the average fidelity of the three
distinct conditional local operations exceeds $2/3$, where
$F_{z4'x6'}$, $F_{x4'z6'}$  are two of these required fidelities and
the third required fidelity is
\begin{eqnarray}
F_{x4'x6'}&=& \frac{1}{4}[P(RR/ + +) + P(LL/ + +) + P(RL/ + -)\notag \\
&&+ P(LR/ + -) + P(RL/ - +) + P(LR/ - +) \notag\\
&&+ P(RR/ - -) + P(LL/ - -)].
 \end{eqnarray}
In our experiment, $F_{x4'x6'}$ is $0.81 \pm 0.02$ calculated from
the data depicted in Fig. 5\emph{C}. The average value of
$F_{z4'x6'}$, $F_{x4'z6'}$ and $F_{x4'x6'}$ is $0.80 \pm 0.01$,
clearly exceeding the boundary of 2/3 and thus proving quantum
parallelism in our gate.

In order to verify the deterministic character of the C-Phase gate,
we have implemented corrections passively  according to different
results of BSMs. Both the qubits $ \left| T \right\rangle_1$ and $
\left| C \right\rangle_2$ are prepared as the state $ \left| +
\right\rangle$. After a C-Phase gate, theoretically the state of
these two qubits should be $\frac{1}{{\sqrt 2 }}(\left| {H+}
\right\rangle + \left| {V-} \right\rangle )$. As depicted in Fig. 6,
without corrections the average fidelity of the output state is only
$0.24 \pm 0.01$. However, with corrections according to the different
results of BSMs, we achieve a state with a fidelity of $0.72 \pm
0.01$. This agrees with the theoretical expectation. In
the future, with the techniques of active feed-forward developed in
\cite{Prevedel}, one can expect to achieve a teleportation-based
deterministic C-Phase gate.

\section{Discussion}

In summary, with two different approaches, we have demonstrated in
principle the feasibility of the GC scheme. By using the six-photon
architecture, we have experimentally realized a C-NOT gate based on
quantum teleportation. The truth table of the gate has been measured
and the ability to entangle two separable qubits has been
demonstrated. With a hyper-entangled four-photon cluster state and
ultra-stable single-photon spatial-polarization BSM, we have
realized and characterized a teleportation-based quantum optical
C-Phase gate.

Below we list some open questions which need to be studied in order
to make an advanced optical system in the future. The off-line
resource states used in the current experiment are equivalent to
four-qubit cluster states. It is proved that efficient preparation
of cluster states is possible with a detector efficiency above 1/2
and an arbitrary small source efficiency \cite{Varnava, duan}, where
an EPR source that emits a vacuum state or a perfect EPR state is
required. This type of state should be studied in the future systems of quantum dots
and ions. Moreover, efforts should also be focused
on the implementation of chip-scale waveguide quantum circuits
\cite{Politi, Matthews}, which can lead to integrated devices.
Third, the spatial modes in the hyper-entangled resource states can
only be connected to the polarization qubit of the same photon with the
current setup. It should be interesting to investigate how to
entangle the spatial mode qubit with the polarization qubit of
another photon \cite{yoran03}. Fourth, the GC scheme  plays an important
role not only in the traditional unitary-evolution-based quantum
computation, but also measurement-based quantum computation. The
demonstration of the GC scheme can be extended to realize universal
``linked-state" measurement-based quantum computation by using more
qubits. Lastly, due to the preparation of the resource states in an
off-line manner and the transversal operating of the gates in GC
scheme, one of the distinct advantages of the GC scheme is that it is
inherently fault-tolerant. Encoding the logic qubits onto an
error-correction code and implementing a fault-tolerant gate
following the GC scheme will be an important step towards
fault-tolerant quantum computation.





\begin{acknowledgments}
We acknowledge Tracy Li for proofreading the manuscript. This work was supported by the National Natural Science Foundation
of China, the Chinese Academy of Sciences and the National
Fundamental Research Program (under Grant No 2006CB921900), the Fundamental Research
Funds for the Central Universities, the European
Commission through the ERC Grant and the STREP project HIP. C.W. was additionally
supported by the Schlieben-Lange Program of the ESF.
\end{acknowledgments}

\section*{Appendix}

\renewcommand{\thefigure}{S\arabic{figure}}
 \setcounter{figure}{0}
\renewcommand{\theequation}{S.\arabic{equation}}
 \setcounter{equation}{0}
 \renewcommand{\thesection}{S.\Roman{section}}
\setcounter{section}{0}
\renewcommand{\thetable}{S.\arabic{table}}

\subsection{Teleportation-based C-NOT gate}
Here, we describe in detail the scheme of a teleportation-based
C-NOT gate. We give a step by step analyses of its implementation
with our setup, shown in Fig. 2 in the main article.

We align each $\beta$-barium borate (BBO) crystal carefully to
produce a pair of polarization entangled photons $i$ and $j$ in the
state:
\begin{equation}
|\Psi^{+}\rangle_{ij} = \frac{1}{\sqrt{2}} \left(|H\rangle_{i}
|H\rangle_{j} + |V\rangle_{i} |V\rangle_{j} \right)
\end{equation}
We use the method described in ref.~\cite{Langford,
Kiesel,Okamoto,Kiesel05} to prepare the cluster state
$|\chi\rangle$. Initially, photons $3$, $4$, $5$ and $6$ are in the
state:
\begin{eqnarray}
|\Psi^{+}\rangle_{34}&\otimes&|\Psi^{+}\rangle_{56}   \nonumber
\\ =&&\frac{1}{2}
\big(|H\rangle_{3}|H\rangle_{4}|H\rangle_{5}|H\rangle_{6} +
|H\rangle_{3}|H\rangle_{4}|V\rangle_{5}|V\rangle_{6} +\nonumber \\
&&|V\rangle_{3}|V\rangle_{4}|H\rangle_{5}|H\rangle_{6} +
|V\rangle_{3}|V\rangle_{4}|V\rangle_{5}|V\rangle_{6} \big).
\end{eqnarray}
We direct photons 4 and 6 to the two input modes of a partially polarizing
beam splitters (PPBS), respectively. The transmission
$T_{H}$ $(T_{V})$ of horizontally (vertically) polarized light at
the PPBS is $1$ $(1/3)$, and we thus get
\begin{eqnarray}
& \rightarrow & \frac{1}{2}
\big(|H\rangle_{3}|H\rangle_{4'}|H\rangle_{5}|H\rangle_{6'} +
\frac{1}{\sqrt{3}}|H\rangle_{3}|H\rangle_{4'}|V\rangle_{5}|V\rangle_{6'} \nonumber \\
& & +\frac{1}{\sqrt{3}}
|V\rangle_{3}|V\rangle_{4'}|H\rangle_{5}|H\rangle_{6'}\nonumber \\
&
&-\frac{1}{3}|V\rangle_{3}|V\rangle_{4'}|V\rangle_{5}|V\rangle_{6'}\big).
\end{eqnarray}
Here we have neglected terms with more than one photon in a single
output mode of the PPBS, since in the experiment we post select only
terms that lead to a six-fold coincidence.

In order to symmetrize the state we place a PPBS' $\left(T_{H} =
1/3, T_{V} = 1\right)$ in each output mode of the PPBS and receive
\begin{eqnarray}
& \rightarrow & \frac{1}{6}
\big(|H\rangle_{3}|H\rangle_{4''}|H\rangle_{5}|H\rangle_{6''} +
|H\rangle_{3}|H\rangle_{4''}|V\rangle_{5}|V\rangle_{6''} \nonumber \\
& & + |V\rangle_{3}|V\rangle_{4''}|H\rangle_{5}|H\rangle_{6''} -
|V\rangle_{3}|V\rangle_{4''}|V\rangle_{5}|V\rangle_{6''}\big).
\end{eqnarray}
This is already the desired four-qubit cluster state up to local
unitary operations. To bring it to the desired form, we place
half-wave plates (HWPs) -- with an angle of $22.5^{\circ}$ between
the fast and the horizontal axis -- into arms 3 and 4. This yields
\begin{eqnarray}
\rightarrow  &&\left(|H\rangle_{3}|H\rangle_{4''} +
|V\rangle_{3}|V\rangle_{4''} \right) |H\rangle_{5}|H\rangle_{6''} \nonumber \\
&& + \left(|H\rangle_{3}|V\rangle_{4''} +
|V\rangle_{3}|H\rangle_{4''} \right) |V\rangle_{5}|V\rangle_{6''}
 \nonumber \\ &=& |\chi\rangle_{34''56''},
\end{eqnarray}
where we have neglected the overall pre-factor 1/6 and we arrive at
the desired ancillary four-photon cluster state $|\chi\rangle$
described in ref.~\cite{gottesman99}.

Photons 1 and 2 constitute the input to our C-NOT gate. We assume
that they are in a most general input state
$|\Psi_{in}\rangle_{12}$, where:
\begin{eqnarray}
|\Psi_{in}\rangle_{ij} &=& \alpha |H\rangle_{i} |H\rangle_{j} +
\beta |H\rangle_{i} |V\rangle_{j}   \nonumber\\
&&+\gamma |V\rangle_{i} |H\rangle_{j} + \delta |V\rangle_{i}
|V\rangle_{j}
\end{eqnarray}
The pre-factors $\alpha$, $\beta$, $\gamma$ and $\delta$ are four
arbitrary complex numbers satisfying $|\alpha|^{2} + |\beta|^{2} +
|\gamma|^{2} + |\delta|^{2} = 1$. Before we proceed, let us define
the desired output state after a C-NOT operation:
\begin{eqnarray}
|\Psi_{out}\rangle_{ij}  & = &  U^{C-NOT}|\Psi_{in}\rangle_{ij} \nonumber\\
& = & \alpha |H\rangle_{i} |H\rangle_{j} + \beta |V\rangle_{i}
|V\rangle_{j} + \nonumber\\
&&\gamma |V\rangle_{i} |H\rangle_{j} + \delta |H\rangle_{i}
|V\rangle_{j}
\end{eqnarray}
The target qubit $i$ is flipped on the condition that the control
qubit $j$ is in the state $|V\rangle$.

We can now express the combined state of all six photons in terms of
Bell states for photons 1\&3 and 2\&5 and in terms of the desired
output state $|\Psi_{out}\rangle_{46}$ for photons 4\&6 with
corresponding Pauli operations:
\begin{widetext}
\begin{align}
\begin{array}{rrcr}
|\Psi_{in}\rangle_{12}\otimes|\chi\rangle_{3456}  = &&&\\
|\Phi^{+}\rangle_{13}|\Phi^{+}\rangle_{25} & |\Psi_{out}\rangle_{46}
& + |\Phi^{+}\rangle_{13}|\Phi^{-}\rangle_{25} &
\hat{\sigma}_{z}^{6}
|\Psi_{out}\rangle_{46} \\
+ |\Phi^{+}\rangle_{13}|\Psi^{+}\rangle_{25} &
\hat{\sigma}_{x}^{4}\hat{\sigma}_{x}^{6}|\Psi_{out}\rangle_{46} & +
|\Phi^{+}\rangle_{13}|\Psi^{-}\rangle_{25} &
\hat{\sigma}_{x}^{4}\hat{\sigma}_{x}^{6}\hat{\sigma}_{z}^{6}
|\Psi_{out}\rangle_{46} \\
+ |\Phi^{-}\rangle_{13}|\Phi^{+}\rangle_{25} &
\hat{\sigma}_{z}^{4}\hat{\sigma}_{z}^{6}|\Psi_{out}\rangle_{46} & +
|\Phi^{-}\rangle_{13}|\Phi^{-}\rangle_{25} & \hat{\sigma}_{z}^{4}
|\Psi_{out}\rangle_{46} \\
+ |\Phi^{-}\rangle_{13}|\Psi^{+}\rangle_{25} &
\hat{\sigma}_{x}^{4}\hat{\sigma}_{z}^{4}\hat{\sigma}_{x}^{6}\hat{\sigma}_{z}^{6}
|\Psi_{out}\rangle_{46} & +
|\Phi^{-}\rangle_{13}|\Psi^{-}\rangle_{25} &
\hat{\sigma}_{x}^{4}\hat{\sigma}_{z}^{4}\hat{\sigma}_{x}^{6}
|\Psi_{out}\rangle_{46}  \\
+ |\Psi^{+}\rangle_{13}|\Phi^{+}\rangle_{25} &
\hat{\sigma}_{x}^{4}|\Psi_{out}\rangle_{46} & +
|\Psi^{+}\rangle_{13}|\Phi^{-}\rangle_{25} &
\hat{\sigma}_{x}^{4}\hat{\sigma}_{z}^{6}
|\Psi_{out}\rangle_{46} \\
+ |\Psi^{+}\rangle_{13}|\Psi^{+}\rangle_{25} &
\hat{\sigma}_{x}^{6}|\Psi_{out}\rangle_{46} & +
|\Psi^{+}\rangle_{13}|\Psi^{-}\rangle_{25} &
\hat{\sigma}_{x}^{6}\hat{\sigma}_{z}^{6}
|\Psi_{out}\rangle_{46}  \\
+ |\Psi^{-}\rangle_{13}|\Phi^{+}\rangle_{25} &
\hat{\sigma}_{x}^{4}\hat{\sigma}_{z}^{4}\hat{\sigma}_{z}^{6}|\Psi_{out}\rangle_{46}
& + |\Psi^{-}\rangle_{13}|\Phi^{-}\rangle_{25} &
\hat{\sigma}_{x}^{4}\hat{\sigma}_{z}^{4}
|\Psi_{out}\rangle_{46} \\
+ |\Psi^{-}\rangle_{13}|\Psi^{+}\rangle_{25} &
\hat{\sigma}_{z}^{4}\hat{\sigma}_{x}^{6}\hat{\sigma}_{z}^{6}|\Psi_{out}\rangle_{46}
& + |\Psi^{-}\rangle_{13}|\Psi^{-}\rangle_{25} &
\hat{\sigma}_{z}^{4}\hat{\sigma}_{x}^{6}
|\Psi_{out}\rangle_{46}  \\
\end{array}
\end{align}
\end{widetext}
With the help of polarizing beam splitters, in our experiment we are
able to identify the Bell states $|\Phi^{\pm}\rangle_{13}$ and
$|\Phi^{\pm}\rangle_{25}$, i.e. we project the combined state of
photons $1$, $2$, $3$ and $5$ onto one of the four possibilities
$|\Phi^{\pm}\rangle_{13}|\Phi^{\pm}\rangle_{25}$. We thus have to
consider four different results of the BSMs: \vskip 1.0cm
\begin{center}
\begin{tabular}{|c|r|}
\hline
Result of BSMs    &   Output state  \\
    \hline
$|\Phi^{+}\rangle_{13}|\Phi^{+}\rangle_{25}$ & $|\Psi_{out}\rangle_{46}$\\
$|\Phi^{+}\rangle_{13}|\Phi^{-}\rangle_{25}$ & $\hat{\sigma}^{6}_{z}|\Psi_{out}\rangle_{46}$\\
$|\Phi^{-}\rangle_{13}|\Phi^{+}\rangle_{25}$ & $\hat{\sigma}^{4}_{z}\hat{\sigma}^{6}_{z}|\Psi_{out}\rangle_{46}$\\
$|\Phi^{-}\rangle_{13}|\Phi^{-}\rangle_{25}$ & $\hat{\sigma}^{4}_{z}|\Psi_{out}\rangle_{46}$\\
    \hline
\end{tabular}
\end{center}
\vskip 1.0cm To receive the desired final state of photons $4$ and
$6$, we have to apply corresponding Pauli operations, depending on
the outcome of the BSMs.

\subsection{Teleportation-based C-Phase gate}

Similarly to the last section, in the implementation of a
teleportation-based C-Phase gate, we also need to apply 16 different
Pauli corrections  on the output qubits according to 16 possible
combinations of outcomes of BSMs. We list the required correction
operations in the Table S.1.

\begin{table}[b]
\begin{center}
\begin{tabular}{ccccc}
  \hline\hline
  measurements & $\left|\Phi^+\right\rangle _3$ & $\left|\Phi^-\right\rangle _3$&$\left|\Psi^+\right\rangle _3$ & $\left|\Psi^-\right\rangle
  _3$\\
  \hline\hline
  $\left|\Phi^+\right\rangle _5$ & $I$ & $Z_{4^{\prime}}$ & $X_{4^{\prime}}Z_{6^{\prime}}$ & $iY_{4^{\prime}}Z_{6^{\prime}}$ \\
  $\left|\Phi^-\right\rangle _5$ & $Z_{6^{\prime}}$ & $Z_{4^{\prime}}Z_{6^{\prime}}$ & $X_{4^{\prime}}$ & $iY_{4^{\prime}}$ \\
  $\left|\Psi^+\right\rangle _5$ & $Z_{4^{\prime}}X_{6^{\prime}}$ & $X_{6^{\prime}}$ & $Y_{4^{\prime}}Y_{6^{\prime}}$ & $-iX_{4^{\prime}}Y_{6^{\prime}}$ \\
  $\left|\Psi^-\right\rangle _5$ & $iZ_{4^{\prime}}Y_{6^{\prime}}$ & $iY_{6^{\prime}}$ & $-iY_{4^{\prime}}X_{6^{\prime}}$ & $-X_{4^{\prime}}X_{6^{\prime}}$ \\
      \hline
\end{tabular}\caption{The required correction operations depending on the Bell
state measurement results. We define $\left|\Phi^ \pm\right\rangle
_{i} = \frac{1}{{\sqrt 2 }}(\left| H \right\rangle _i \left| H'
\right\rangle _i  \pm \left| V \right\rangle _i \left| V'
\right\rangle _i ), \left|\Psi^ \pm \right\rangle_i   =
\frac{1}{{\sqrt 2 }}(\left| H \right\rangle _i \left| V'
\right\rangle _i  \pm \left| V \right\rangle _i \left| H'
\right\rangle _i )$, where $\left| H' \right\rangle$ and $\left| V'
\right\rangle$ are spatial qubits. The 16 cases in Fig.~6 in the
main text correspond to the output results:
$\left|\Phi^+\right\rangle _3 \left|\Phi^+\right\rangle _5$,
$\left|\Psi^+\right\rangle _3 \left|\Phi^+\right\rangle _5$,
$\left|\Phi^+\right\rangle _3 \left|\Psi^+\right\rangle _5$,
$\left|\Psi^+\right\rangle _3 \left|\Psi^+\right\rangle _5$,
$\left|\Phi^+\right\rangle _3 \left|\Phi^-\right\rangle _5$,
$\left|\Psi^+\right\rangle _3 \left|\Phi^-\right\rangle _5$,
$\left|\Phi^+\right\rangle _3 \left|\Psi^-\right\rangle _5$,
$\left|\Psi^+\right\rangle _3 \left|\Psi^-\right\rangle _5$,
$\left|\Phi^-\right\rangle _3 \left|\Phi^-\right\rangle _5$,
$\left|\Psi^-\right\rangle _3 \left|\Phi^-\right\rangle _5$,
$\left|\Phi^-\right\rangle _3 \left|\Psi^-\right\rangle _5$,
$\left|\Psi^-\right\rangle _3 \left|\Psi^-\right\rangle _5$,
$\left|\Phi^-\right\rangle _3 \left|\Phi^+\right\rangle _5$,
$\left|\Psi^-\right\rangle _3 \left|\Phi^+\right\rangle _5$,
$\left|\Phi^-\right\rangle _3 \left|\Psi^+\right\rangle _5$,
$\left|\Psi^-\right\rangle _3 \left|\Psi^+\right\rangle _5$.  }
\end{center}
\end{table}


\end{thebibliography}

\end{document}